\documentstyle[a4,12pt]{article}
\renewcommand{\baselinestretch}{1.55}

\begin{document}
\parskip=5pt plus 1pt minus 1pt

\def\dis{\displaystyle}

\begin{flushright}
{\large\bf LMU-21/94} \\
{November 1994}
\end{flushright}

\vspace{0.5cm}

\begin{center}
{\Large\bf Wolfenstein Parametrization Reexamined}
\end{center}

\vspace{0.8cm}

\begin{center}
{\bf Zhi-zhong Xing}\footnote{E-mail: Xing$@$hep.physik.uni-muenchen.de}
\end{center}

\begin{center}
{\sl Sektion Physik, Ludwig-Maximilians-Universit${\sl\ddot a}$t M${\sl\ddot
u}$nchen}\\
{\sl Theresienstra$\beta$e 37, D-80333 Munich, Germany}
\end{center}

\vspace{1.5cm}

\begin{abstract}

The Wolfenstein parametrization of the $3\times 3$ Kobayashi-Maskawa (KM)
matrix $V$ is modified by keeping its unitarity up to the accuracy of
$O(\lambda^{6})$.
This modification can self-consistently lead to the off-diagonal asymmetry of
$V$: $|V_{ij}|^{2}-|V_{ji}|^{2}$ = $Z\displaystyle\sum_{k}\epsilon^{~}_{ijk}$
with $Z\approx A^{2}\lambda^{6} (1-2\rho)$, which is comparable in magnitude
with the
Jarlskog parameter of $CP$ violation $J\approx A^{2}\lambda^{6}\eta$ . We
constrain
the ranges of $J$ and $Z$ by using the current experimental data, and point out
that
the possibility of a symmetric KM matrix has almost been ruled out.

\end{abstract}

\vspace{1.5cm}

\begin{center}
\end{center}

\newpage

Within the standard electroweak model, the $3\times 3$ Kobayashi-Maskawa (KM)
matrix $V$ offers a natural description of quark mixing and $CP$ violation [1].
The unitarity of $V$ leads to a rephasing invariant measure of $CP$
violation (the Jarlskog parameter $J$ [2])
\begin{equation}
{\rm Im}\left (V_{il}V_{jm}V^{*}_{im}V^{*}_{jl} \right )
\; =\; J\sum_{k,n}\epsilon^{~}_{ijk}\epsilon^{~}_{lmn}
\end{equation}
and an off-diagonal asymmetry of $V$ (denoted by $Z$)
\begin{equation}
|V_{ij}|^{2}-|V_{ji}|^{2} \; =\; Z\sum_{k}\epsilon^{~}_{ijk} \; ,
\end{equation}
where $i,j,k,l,m,n=1,2,3$. Confronting these two relations with
the existing and forthcoming experimental data may provide
a stringent test of the standard model.

It proves convenient in practice to use a parametrization form of the KM
matrix [1,3,4]. Among the proposed parametrizations, the Wolfenstein form [3]
\small
\begin{equation}
V_{\rm W} \; =\; \left (
\begin{array}{ccc}
1-\frac{1}{2}\lambda^{2} & \lambda & A\lambda^{3}\left [\rho-i\eta \left
(1-\frac{1}{2}\lambda^{2}\right )\right ] \\
-\lambda	& 1-\frac{1}{2}\lambda^{2}-iA^{2}\lambda^{4}\eta & A\lambda^{2}\left
(1+i\lambda^{2}\eta\right ) \\
A\lambda^{3}\left (1-\rho-i\eta\right )	& -A\lambda^{2}	& 1
\end{array}
\right )
\end{equation}
\normalsize
is very popular for phenomenological applications. However, there are two minor
drawbacks
associated with $V_{\rm W}$: (a) its unitarity is only kept up to the accuracy
of $O(\lambda^{4})$;
and (b) it cannot self-consistently describe the off-diagonal asymmetry $Z$.
These are of course
unsatisfactory when we apply $V_{\rm W}$ to more precise experimental data of
quark mixing
and $CP$ violation.
Noticing drawback (a), Kobayashi [5] has recently presented an exactly unitary
parametrization of
the KM matrix $V$ in terms of the Wolfenstein parameters. The exactness
of this parametrization, accompanied with a complicated form, reduces its
phenomenological
practicability on the other hand.

Following the same approach as that of Kobayashi and keeping unitarity up to
the
accuracy of $O(\lambda^{6})$, here we present a
modified version of the Wolfenstein parametrization as follows:
\small
\begin{equation}
V^{'}_{\rm W} = \left (
\begin{array}{ccc}
1-\frac{1}{2}\lambda^{2}-\frac{1}{8}\lambda^{4}	& \lambda	& A\lambda^{3}(\rho
-i\eta) \\
-\lambda \left [1+ \frac{1}{2}A^{2}\lambda^{4}(2\rho -1) + i
A^{2}\lambda^{4}\eta \right ]	&
1-\frac{1}{2}\lambda^{2}-\frac{1}{8}\left (4A^{2}+1\right)\lambda^{4} 	&
A\lambda^{2} \\
A\lambda^{3}(1-\rho-i\eta)	& -A\lambda^{2}\left [1+\frac{1}{2}\lambda^{2}(2\rho
-1 )
+i\lambda^{2}\eta \right ] 	& 1-\frac{1}{2}A^{2}\lambda^{4}
\end{array}
\right ) \; .
\end{equation}
\normalsize
One can observe a few different features of $V^{'}_{\rm W}$ from $V_{\rm W}$:

(1) The matrix elements $|V_{12}|$ and $|V_{21}$ given by $V^{'}_{\rm W}$ are
not symmetric.

(2) $|V_{22}|$ is smaller than $|V_{11}|$ and the difference is of the order
$A^{2}\lambda^{4}/2$.
This is in agreement with the prediction from a variety of quark mass
$Ans\ddot{a}tze$ [6].

(3) The relation between $V_{ub}$ and $V_{cb}$, described by $\lambda$, $\rho$,
and $\eta$,
becomes simpler in $V^{'}_{\rm W}$ . Thus it should be more convenient to
confront the
ratio $V_{ub}/V_{cb}$ with the data of $B$-meson physics and $CP$ violation
[7].

(4) Both the normalization conditions and orthogonality relations of
$V^{'}_{\rm W}$
can be given to the degree of accuracy $O(\lambda^{6})$.

With the help of $V^{'}_{\rm W}$, we are now able to carry out a
self-consistent calculation of
$Z$ (as well as $J$) and obtain
\begin{equation}
Z \; \approx \; A^{2}\lambda^{6} (1-2\rho) \; , \;\;\;\;\;\;\;\;\;\;\;\;
J \; \approx \; A^{2}\lambda^{6} \eta \; .
\end{equation}
Clearly $Z$ is of the same order as $J$. Note that $Z$ is approximately
independent of $\eta$,
a parameter necessary for $CP$ violation. In the case of $\rho=1/2$,
$Z\approx 0$ holds, which implies a symmetric quark mixing matrix $V$ up to the
accuracy of $O(\lambda^{6})$. The experimental value of $\rho$ lies in the
range $-0.6\leq \rho\leq 0.5$, but it is most likely around zero [8]. Thus the
possibility of a symmetric KM matrix $V$ has almost been ruled out [9]. In Fig.
1 we
plot the ranges of $J$ and $Z$ allowed by the current data on $A, \lambda,
\rho$, and
$\eta$ . The linear relation between $J$ and $Z$, given by
\begin{equation}
Z\; \approx \; \frac{1-2\rho}{\eta}J \; ,
\end{equation}
can be clearly observed from Fig. 1. We expect that Eq. (6) could serve as a
good test of
unitarity of the KM matrix in the near future.

In summary, a modified form of the Wolfenstein
parametrization has been presented, in which unitarity is kept up to the
accuracy of $O(\lambda^{6})$.
With this new parametrization we have carried out a self-consistent
calculation of the off-diagonal asymmetry of $V$ and found that it is
comparable in magnitude
with the rephasing invariant measure of $CP$ violation. The constraints on
these two parameters
are given by using the current experimental data. We conclude
that the possibility of a symmetric KM matrix has almost been ruled out. \\

	The author would like to thank Professor H. Fritzsch for his kind
hospitality and constant encouragements. He is also indebted to the
Alexander von Humboldt Foundation for its financial support.


\newpage

\begin{figure}
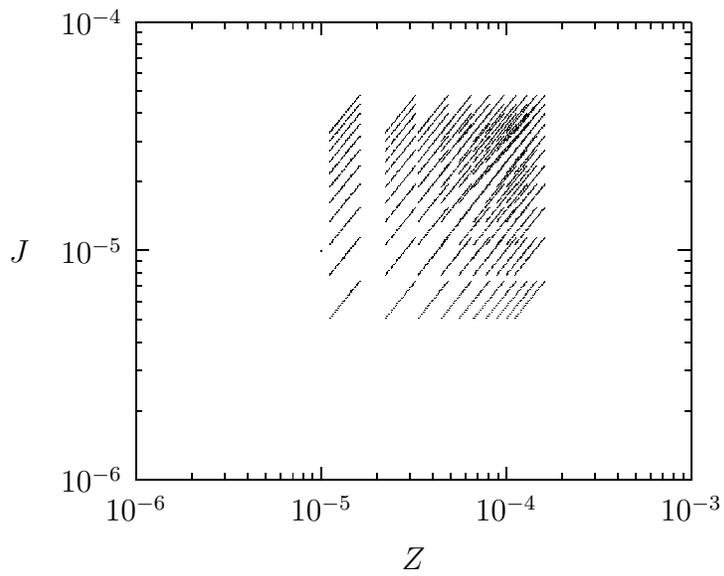

\setlength{\unitlength}{0.240900pt}
\ifx\plotpoint\undefined\newsavebox{\plotpoint}\fi

\caption{The ranges of $J$ and $Z$ allowed by the current experimental data on
$A, \lambda, \rho$, and $\eta$ .}
\end{figure}


\newpage

\begin{thebibliography}{99}
\bibitem{1} M. Kobayashi and T. Maskawa, Prog. Theor. Phys. 49 (1973) 552.
\bibitem{2} C. Jarlskog, Phys. Rev. Lett. 55 (1985) 1039; \\
D. D. Wu, Phys. Rev. D33 (1986) 860; \\
I. Dunietz, O. W. Greenberg, and D. D. Wu, Phys. Rev. Lett. 55 (1985) 2935.
\bibitem{3} L. Wolfenstein, Phys. Rev. Lett. 51 (1983) 1945.
\bibitem{4} H. Fritzsch and J. Plankl, Phys. Rev. D35 (1987) 1732; \\
H. Harari and M. Leurer, Phys. Lett. B181 (1986) 123; \\
H. Fritzsch, Phys. Rev. D32 (1985) 3058; \\
L. L. Chau and W. Y. Keung, Phys. Rev. Lett. 53 (1984) 1802; \\
L. Maiani, Phys. Lett. B62 (1976) 183.
\bibitem{5} M. Kobayashi, Prog. Theor. Phys. 92 (1994) 287; 92 (1994) 289.
\bibitem{6} H. Fritzsch and Z. Z. Xing, in preparation.
\bibitem{7} A. J. Buras, M. E. Lautenbacher, and G. Ostermaier, Phys. Rev. D50
(1994) 3433.
\bibitem{8} Particle Data Group, M. Aguilar-Benitez et al., Phys. Rev. D50
(1994) 1173; \\
A. Ali and D. London, Report No. CERN-TH.7398/94 (to appear in Z. Phys. C); \\
A. J. Buras, Phys. Lett. B333 (1994) 476.
\bibitem{9} A few symmetric KM $Ans\ddot{a}tze$ and their phenomenological
implications
have been studied in the literature. See, e.g., \\
P. Kielanowski, Phys. Rev. Lett. 63 (1989) 2189; \\
J. L. Rosner, Phys. Rev. Lett. 64 (1990) 2590; \\
G. C. Branco and P. A. Parada, Phys. Rev. D44 (1991) 923; \\
M. Tanimoto, Mod. Phys. Lett. A6 (1991) 2309; \\
G. Belanger, C. Hamzaoui, and Y. Koide, Phys. Rev. D45 (1992) 4186; \\
Y. Nir and U. Sarid, Phys. Rev. D47 (1993) 2818.

\end{thebibliography}
\end{document}